\documentclass[11pt]{article}
\usepackage{hyperref}
\pdfoutput=1

\begin{document}

\title{Surfactant-assisted spreading\\ of an oil-in-water emulsion \\on a liquid bath }

\author{M. Roch\'{e}\dag, Z. Li\dag, I. Griffiths\dag\\ A. Saint-Jalmes\ddag, H.A. Stone\dag \\
\\ \vspace{2pt} \\ \tiny{\dag Complex Fluids Group, Department of Mechanical and Aerospace Engineering}\\
\tiny{Princeton University, 08542 Princeton NJ, USA}\\ \tiny{\ddag Institut de Physique de Rennes, Universit\'{e} Rennes 1}\\ \tiny{263 avenue du G\'{e}n\'{e}ral Leclerc, 35700 Rennes, France}}

\maketitle

\begin{abstract}
This fluid dynamics video shows how an oil-in-water emulsion stabilized by an ionic surfactant spreads on the free surface of a layer of pure water. The spreading shows two intriguing features: a transparent area surrounding the source of oil droplets, and a fast retraction of the layer of oil droplets on itself once the source has emptied. We show that the dynamics of spreading are strongly connected to the interfacial/bulk properties of the surfactant.
\end{abstract}

The \href{./anc/SurfactantAssistedSpreading_emulsions_web.mpg}{video} depicts the spreading behavior of an olive oil-in-water emulsion stabilized by an ionic surfactant on the free surface of a layer of pure water. Cetyltrimethylammonium bromide (CTAB), decyltrimethylammonium bromide (DeTAB) and sodium dodecyl sulfate (SDS) were used as the surfactants at a concentration of 20 times their critical micellar concentration (CMC; $CMC_{SDS}=2.3 g/L$, $CMC_{CTAB}=0.353g/L$, $CMC_{DeTAB}=18.22g/L$). We report two stunning features of these systems: a transparent area surrounding the source of oil droplets, and a fast retraction of the layer of oil droplets on itself once the source has emptied. We show using the different surfactants that the dynamics of spreading are strongly connected to their interfacial/bulk properties.
\end{document}